\documentclass[prl,twocolumn,showpacs,preprintnumbers,floatfix]{revtex4-1}
\usepackage{graphicx}
\usepackage{dcolumn}
\usepackage{bm}
\usepackage{latexsym,epsfig}
\usepackage{graphicx}
\usepackage{verbatim}
\usepackage{comment}
\usepackage{amsmath}
\usepackage{amssymb}
\usepackage{stmaryrd}
\usepackage{color}

\newcommand{\bfig}{\begin{figure}}
\newcommand{\efig}{\end{figure}}

\begin{document}
\title{Anomalous pairing of bosons: Effect of multi body interactions in optical lattice}

\author{Manpreet Singh$^{1}$, Sebastian Greschner$^{2,3}$ and Tapan Mishra$^{1}$}
\affiliation{$^{1}$Department of Physics, Indian Institute of Technology, Guwahati, Assam - 781039, India}
\affiliation{$^{2}$Institut f\"ur Theoretische Physik, Leibniz Universit\"at Hannover, 30167~Hannover, Germany}
\affiliation{$^{3}$Department of Quantum Matter Physics, University of Geneva, 1211 Geneva, Switzerland}

\date{\today}

\begin{abstract}
An interesting first order type phase transition between Mott lobes has been reported in Phys. Rev. Lett. {\bf 109}, 135302 (2012) for 
a two-dimensional Bose-Hubbard model in the presence of attractive three-body interaction. We re-visit the scenario in a system 
of ultracold bosons in a one-dimensional optical lattice using the density matrix renormalization group method and show that an 
unconventional pairing of particles occurs due to the competing two-body repulsive and three-body attractive interactions. This 
leads to a pair superfluid phase sandwiched between the Mott insulator lobes corresponding to densities $\rho=1$ and 
$\rho=3$ in the strongly interacting regime. We further extend our analysis to a 
two dimensional Bose-Hubbard model using the self consistent cluster-mean-field theory approach and confirm that the unconventional 
pair superfluid phase stabilizes in the region between the Mott lobes in contrast to the direct first order jump as predicted before. 
In the end we establish connection to the most general Bose-Hubbard model and 
analyse the fate of the pair superfluid phase in presence of an external trapping potential.
\end{abstract}


\maketitle

Ultracold atoms in optical lattices have lead to numerous exotic phenomena in recent years. An exquisite control over 
the system parameters such as the tunneling and interaction among atoms leads to the path breaking observation of the 
superfluid (SF)-Mott insulator (MI) transition in optical lattice followed by its theoretical prediction~\cite{Bloch2,Zoller}. 
This is an interesting manifestation 
of the many-body physics of simple Bose-Hubbard model. A strong on-site two-body repulsion which can be controlled by 
the technique of Feshbach resonance or by tuning the laser intensity, freezes the particle motion in the lattice and 
drives the system from the SF phase to MI phase. In addition to the two-body interaction, ultracold atomic systems in 
optical lattices also possess on-site three- and higher body interactions whose natural existence has been found in 
recent experiment~\cite{SWill}. Using the Feshbach resonance technique it is also possible, in principle, to tune the 
scattering length in such a way that the atomic interactions can become attractive in nature. It is to be noted that a 
small attractive interaction in an optical lattice leads to the collapse of all the atoms onto a single 
site~\cite{Dalfovo}. In such a scenario a tunable three or higher body interaction stabilises the system against 
collapse, e.g. a very strong three-body repulsion could prevent more than two atoms per site and an ensemble of pairs 
of bosons can be created. Under proper conditions of density and two-body interactions, the system exhibits a pair 
superfluid (PSF) phase which can be characterised by the finite off diagonal long range order (ODLRO) of composite 
particles or pairs. The infinitely strong three-body interaction can arise as a result of three-body loss process due 
to the elastic scattering of atoms~\cite{Baranov}. 

Moreover, proposals have been made to engineer such three-body as well as higher order interactions in bosonic lattice 
systems in a controlled manner~\cite{Petrov1,Petrov2,Daley2}. The effect of three-body interaction on the SF-MI 
transition has been studied extensively in the context of Bose-Hubbard model. As mentioned before, a critical two-body 
attraction between atoms and three-body hardcore constraint may lead to the PSF-SF (henceforth called atomic superfluid (ASF))
transition~\cite{Baranov,Wessel,MSingh1,YZhang}. On the contrary, for two-body repulsive interaction, the MI lobe for 
$\rho>1$ expands (shrinks) for repulsive (attractive) three-body interaction~\cite{TSowinski1,MSingh1}. It is to be 
noted that the large three-body attraction will eventually lead to collapse. Hence, a strong four-body repulsion is
required which can stabilize the system against such a collapse~\cite{TSowinski1}.

Recently, in an interesting proposal, it was shown that the three-body interaction can be engineered in such a way that 
it affects only the triply occupied sites~\cite{Sansone}. This special form of the three-body interaction which is 
different from the conventional three-body interaction modifies the Bose-Hubbard model as;

\begin{eqnarray}
H=&-&t\sum_{<i,j>}(a_{i}^{\dagger} a_{j}+H.c.)\nonumber\\
&+&{{U}\over{2}}\sum_{i}n_i(n_i-1)+ W\sum_i \delta_{n_i,3}
\label{eq:zero}
\end{eqnarray}  
where ${a_i}^{\dagger}({a_i})$ is the bosonic creation (annihilation) operator, $n_i$ is the number operator for the $i^{th}$ site,
and $i,~j$ are site indices. $U$ and $W$ are the on-site two- and three-body interactions and $t$ is the hopping amplitude between 
the nearest neighbour sites $<i,j>$.
 
\begin{figure}[!t]
   \centering
\includegraphics[width=0.4\textwidth]{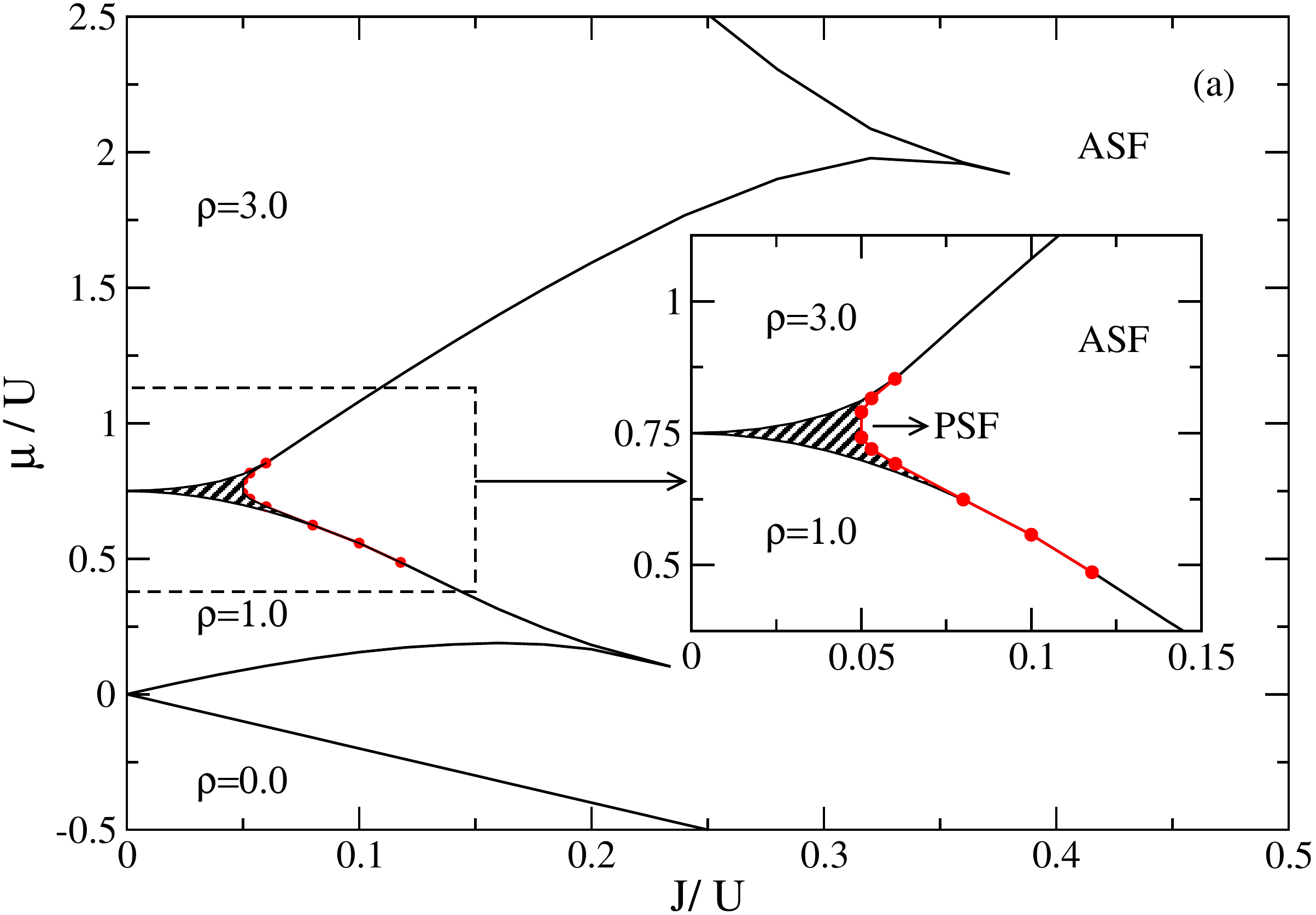}
\includegraphics[trim={0.3cm 16.0cm 5.5cm 0.0cm},clip,width=0.5\textwidth]{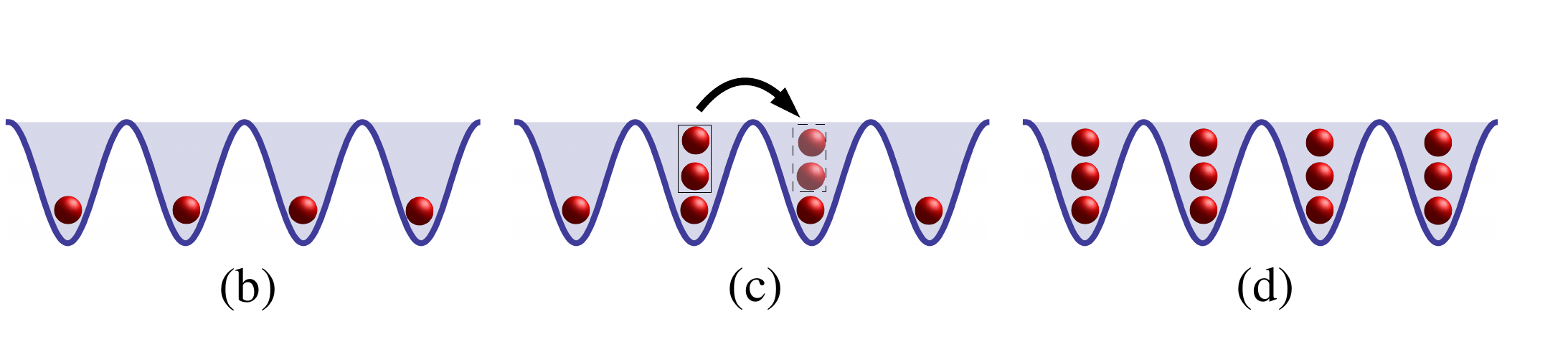}
\vspace{-0.8cm}
\caption{(Color online) (a) Phase diagram for the model given in Eq.~\ref{eq:zero}
for $W=-1.5 U$, obtained using DMRG.
Inset shows the zoomed in part of the main phase diagram which is enclosed in a dashed-line box, 
(b) MI phase at $\rho=1.0$, (c) PSF phase, and (d) MI phase at $\rho=3.0$.}
\label{fig:PD_DMRG}
\vspace{-0.4cm}
\end{figure}

It was shown that the model shown in Eq.~\ref{eq:zero} exhibits shrinking of the MI lobes for $\rho=2$ and $\rho=4$ 
(henceforth denoted as MI(2) and MI(4)) as a function of repulsive $U$ and 
attractive $W$ in a two-dimensional($2d$) Bose-Hubbard model eventually leading to a first-order transition from the MI(1) to MI(3) 
lobe~\cite{Sansone}. This first order transition was attributed to the suppression of double occupancy in favour of triply occupied 
sites in the limit of dominant three-body attraction. In this limit the doping of particles (holes) in MI(1) (MI(3)) lobe plays very 
important role.
In such a scenario we ask a question, whether a similar phenomena can occur in one dimension where correlation effects are 
maximum?

To address this question, we revisit the Bose-Hubbard model shown in Eq.~\ref{eq:zero} for a one dimensional optical lattice using the 
density matrix renormalization group (DMRG) method~\cite{white}. By considering repulsive $U$ and attractive $W$, and for the specific 
value of $W=-1.5U$, we obtain the complete phase diagram. Interestingly, in this case we see a counter intuitive situation where bosons pair up 
to form a pair superfluid phase (PSF). This anomalous PSF phase is sandwiched between the MI(1) and MI(3) lobes for large values of 
interaction strengths as shown in the DMRG phase diagram in Fig.~\ref{fig:PD_DMRG}(a). In our DMRG simulations we use system sizes up to 
$80$ sites and $800$ density matrix eigen states by keeping $n_{max}=6$. We discuss about the details of the phases and phase transitions 
shown in the phase diagram of Fig.~\ref{fig:PD_DMRG}(a), in the following paragraphs. 

\begin{figure}[!t]
\centering
\includegraphics[width=0.4\textwidth]{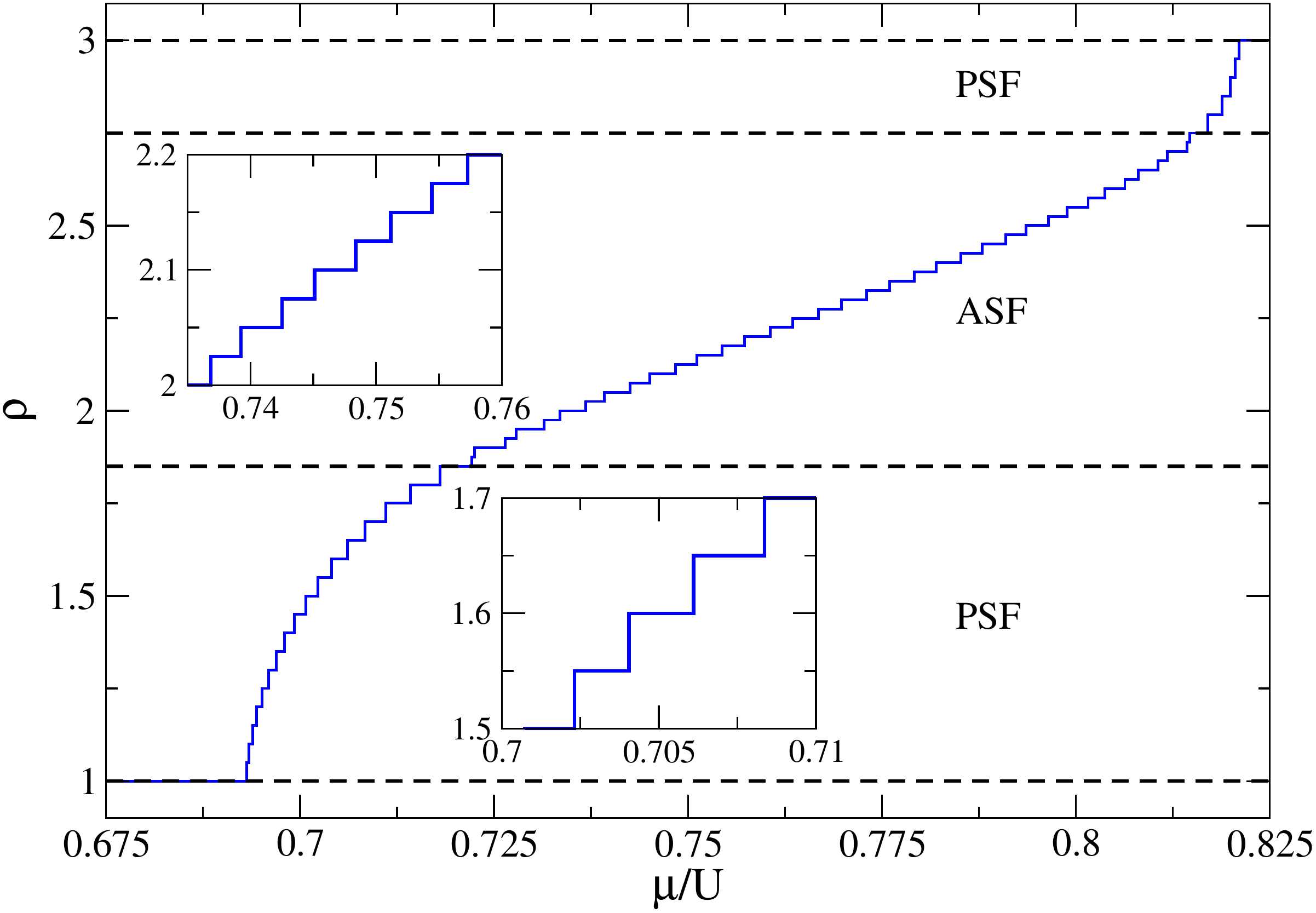}
\vspace{-0.1cm}
\caption{Equation of state $\rho$ vs $\mu$ for $W=-1.5U$ and $t/U=0.053$. 
In the PSF regions the finite system size data exhibits jumps of $\Delta N=2$, 
while the ASF phase is characterised by $\Delta N=1$.}
\label{fig:dmrg_mu_rho}
\vspace{-0.3cm}
\end{figure}
 


The phase diagram of Fig.~\ref{fig:PD_DMRG}(a) shows different quantum phases, namely, the MI(1), MI(3), the ASF and the PSF phases.
We fix the ratio $W=-1.5U$ and vary $U$ to obtain the complete phase diagram by using the signatures from the chemical potential 
($\mu$) vs. density ($\rho$) from our DMRG calculation~\cite {TMishra1}. The gapped MI phases are characterised by the plateaus in 
the $\rho$ vs. $\mu$ plot, whereas, the superfluid phases appear around the Mott plateaus. The areas bounded by the black lines 
are the MI lobes, the shaded region bounded by red dots between the two MI lobes is the PSF phase and the remaining space is the ASF 
phase in the phase diagram in Fig.\ref{fig:PD_DMRG}(a). The existence of the PSF phase can be seen from the $\rho$ vs. $\mu$ curve 
as shown in Fig.~\ref{fig:dmrg_mu_rho} for a cut through the phase diagram at $t/U=0.053$. 
It can be clearly seen that the MI(1) and MI(3) plateaus appear at $\rho=1$ and $\rho=3$. In the region between the two MI phases the 
density jumps in discrete steps with respect to the chemical potential which is a signature of the superfluid phase. Within the superfluid 
region the PSF phase can be discerned by the characteristic finite system size behaviour which exhibits a 
series of jumps of $\Delta N=2$, whereas, the jump of $\Delta N=1$ particle indicates the ASF phase. Here $N$ is the total particle 
number of the system. The insets of 
Fig.~\ref{fig:dmrg_mu_rho} show a zoomed in section of PSF and ASF regions.

\begin{figure}[t]
\centering
\includegraphics[width=0.4\textwidth]{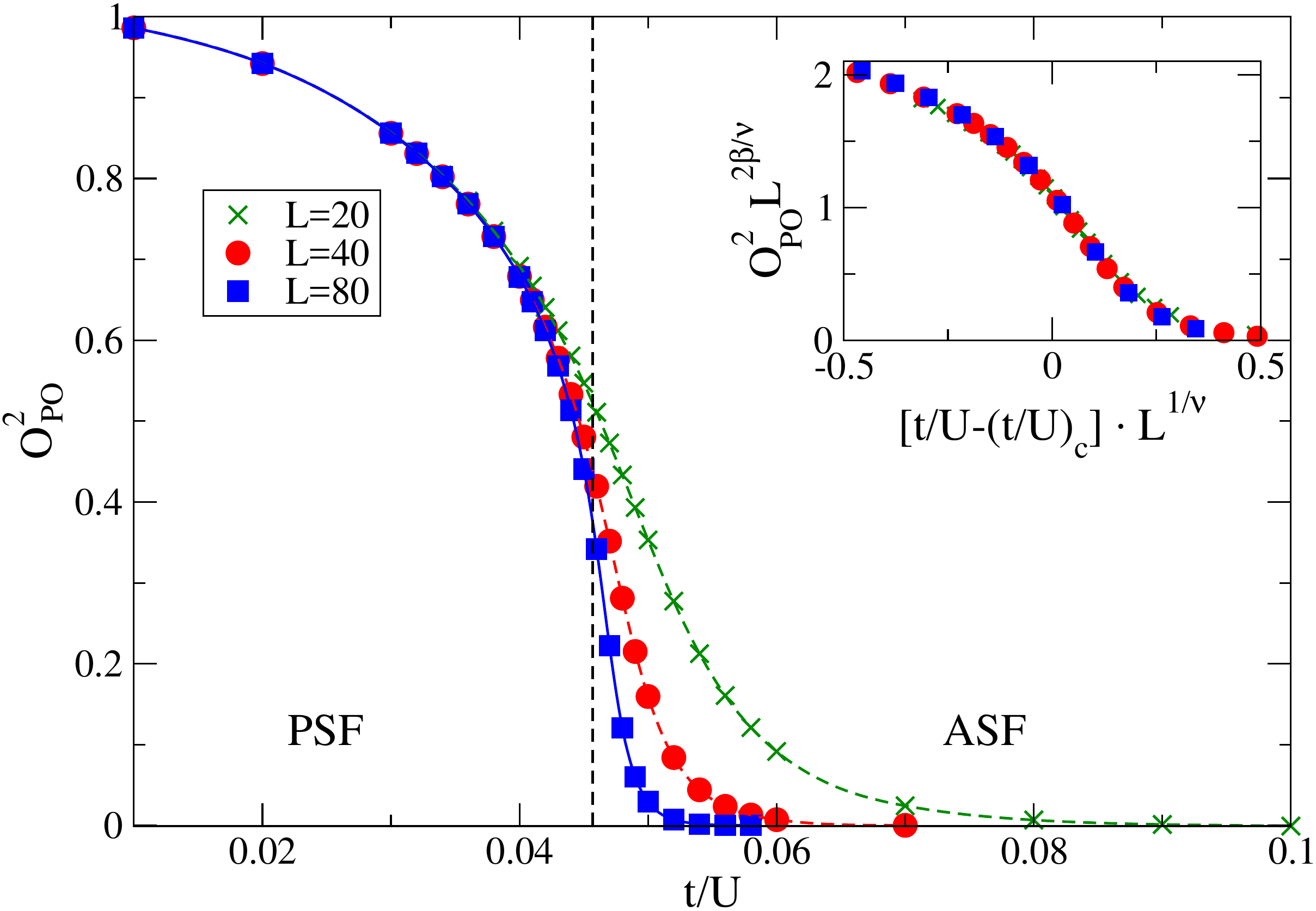}
\caption{
Parity order parameter for the PSF to ASF transition as function of $(t/U)$ using various system 
sizes $L$ for $W=-1.5U$ and $\rho=2$. The inset shows the collapse of all 
finite size data to one curve close to the phase transition point $(t/U)_c \approx 0.045 U$ according to 
typical Ising scaling relations.}
\label{fig:pofs}
\vspace{-0.3cm}
\end{figure}

In order to further understand this pairing phenomena, we analyze the parity order which has been measured in the ultracold quantum 
gas experiments recently~\cite{Bakr,Endres}. The parity order parameter is defined as 
\begin{align}
O_P^2 (i,j) = \langle {\rm e}^{i \sum_{i<k<j} \pi n_{k}} \rangle \;.
\end{align}
Here $n_k$ stands for the number operator at $k^{th}$ site. 
In Fig.~\ref{fig:pofs} we plot the parity order $O_P^2 \equiv O_P^2 (L/4,3L/4)$ as a function of $(t/U)$ for 
system sizes of $L=20, 40, 80$ at $\rho=2$. It is clear from the figure that the value of $O_P^2$ is finite in the PSF 
phase and gradually vanishes in the ASF phase as the system size increases and the point of vanishing parity order parameter 
approaches the critical value $(t/U)_c\approx 0.045$ for the PSF-ASF transition. This PSF-ASF transition is found to be of Ising 
universality class from the finite size scaling of the data. We plot $L^{2\beta/\nu}O_P^2$ with respect to 
$[t/U-(t/U)_c]L^{1/\nu}$ in the inset of Fig.~\ref{fig:pofs} for different lengths. Here $\beta$ represents the critical exponent 
corresponding to the universality class of the phase transition and $\nu$ is the correlation exponent. The complete collapse of the 
data for $(t/U)_c\approx0.045$, $\beta=1/8$ and $\nu=1$ shows the Ising type PSF-ASF transition. The Ising nature of this transition 
is further complemented by calculating the fidelity susceptibility~\cite{Gu} which shows a divergent peak as one approaches the 
transition point (see supplementary material for details).

Although it seems intriguing to have a PSF phase in such a model with repulsive two-body interaction, one can understand this by 
analyzing the possible doping of particle (holes) from the MI(1) (MI(3)) phases in the region between the lobes. In general, when the 
system is in the Mott phases, any allowed doping should be in steps of single particle or hole for a purely repulsive Bose-Hubbard 
model. However, the presence of strong attractive $W$ favors an on-site three-particle state. As a result, when the chemical 
potential $\mu$ increases gradually, the system accumulates two particles at a time to minimize the energy. Eventually, the system 
enters into the MI(3) phase at $\rho=3$ skipping the MI(2) phase. 
As the uniform MI(1) phase in the background provides a constant repulsive interaction, a pair of particles can always hop around the 
lattice without any cost of energy. This process of two particles getting added at a time continues till the system reaches the MI(3) 
phase and these pairs of particles leads to the PSF phase. This phenomenon is depicted in Fig.~\ref{fig:PD_DMRG}(a, b and c). 
The MI(1)-PSF and PSF-MI(3) transitions are continuous phase transitions. 
This is confirmed from the absence of the discontinuous jump in the $\rho$ vs. $\mu$ curve. We also find that the MI(1)-ASF and 
ASF-MI(3) phase transitions are continuous in nature for the entire parameter space considered. 

\begin{figure}[!t]
   \centering
\includegraphics[width=0.4\textwidth]{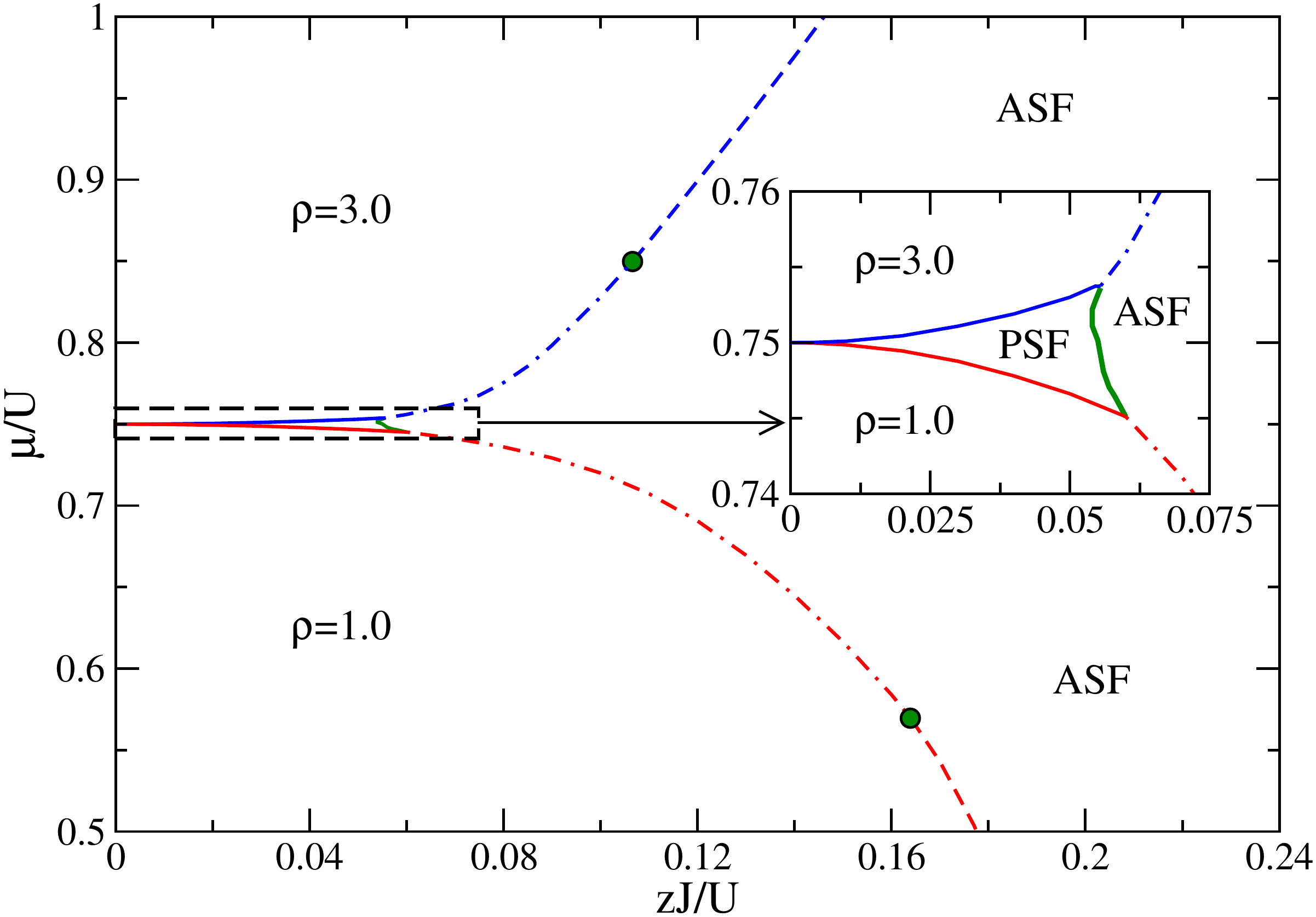}
\caption{(Color online) Phase diagram for the model given in Eq.\ref{eq:zero} for $W=-1.5U$, obtained using 2d-CMFT.
Dashed lines and dot-dashed lines indicate the first-order and second order phase transition boundaries respectively.
Green circles indicate the tri-critical points where the first and second order phase boundaries meet. The region between the MI lobes
bounded by the solid green line shows the PSF phase. 
Inset shows the a zoomed in part of the main phase diagram for clarity.}
\label{fig:PD_CMFT}
\vspace{-0.3cm}
\end{figure}

After getting the insight about the PSF phase in between the MI lobes for $1d$ chain, one may ask why the above mentioned argument of 
boson pairing will not hold for a two dimensional optical lattice system ? To this end we re-investigate the two-dimensional 
Bose-Hubbard model for $W=-1.5U$, using the self consistent cluster mean field theory (CMFT) 
approach~\cite{McIntosh} in the grand canonical ensemble. CMFT works very well for the simple 
models like Eq.~\ref{eq:zero} and matches fairly well with the Quantum Monte Carlo results. By using a four sites cluster and analyzing 
the ground state properties we obtain the complete phase diagram as shown in Fig.\ref{fig:PD_CMFT}. The MI(1) and MI(3) lobes are 
demarcated by the red and blue boundaries. 
Interestingly, in this case also we obtain a finite region of PSF phase which survives between the MI(1) and MI(3) phases bounded by the 
solid green line. This result is in contrast with the direct first order MI(1)-MI(3) phase transition shown in Ref.~\cite{Sansone}. 
The inset of Fig.~\ref{fig:PD_CMFT} shows the zoomed in region depicting the PSF phase.

\begin{figure}[!t]
   \centering
\includegraphics[width=0.45\textwidth]{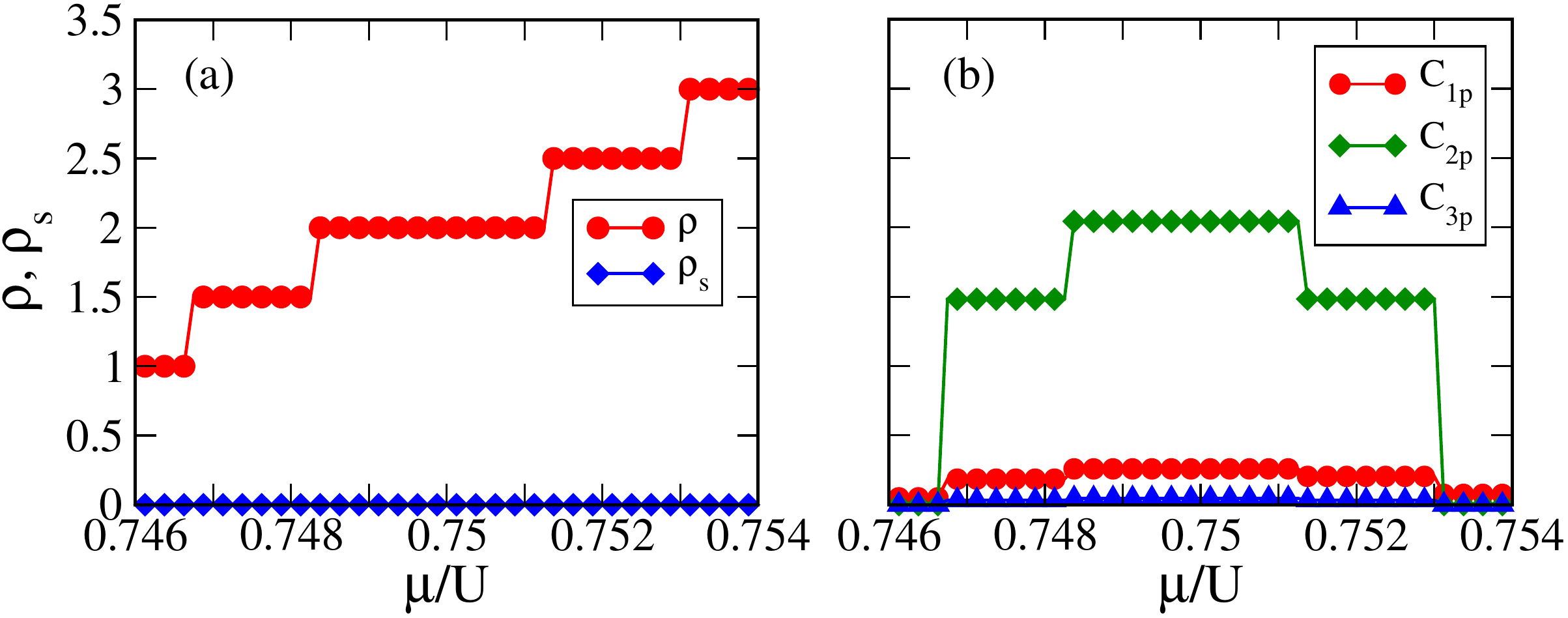}
\caption{(Color online) (a) $\rho,\rho_s-\mu$ plot corresponding to a cut along $zt/U=0.05$ of Fig.~\ref{fig:PD_CMFT}.
(b) Corresponding one-, two- and three-particle correlation functions ($C_{1p}$, $C_{2p}$ and $C_{3p}$) 
showing the clear signature of the PSF phase(see text).}
\label{fig:corr_CMFT}
\vspace{-0.3cm}
\end{figure}

This phase diagram is obtained by using the behavior of density $\rho$ and the superfluid density $\rho_s=\phi^2$(where $\phi$ is the 
superfluid order parameter) with respect to the chemical 
potential $\mu$. We plot $\rho$ and $\rho_s$ with respect to $\mu$ along a cut through the CMFT phase diagram of Fig.~\ref{fig:PD_CMFT}
for $zt/U=0.05$ which passes through the MI(1), PSF and MI(3) phases.  
The discrete jumps in the $\rho$ vs. $\mu$ plot(red circles) in steps of $\Delta N=2$ indicates the 
PSF phase as shown in Fig.\ref{fig:corr_CMFT}(a). The plateaus at $\rho=1$ and $\rho=3$ are due to the MI phases. The value of $\rho_s$ 
is zero throughout the cut which is shown as the blue diamonds. 
Further in Fig.\ref{fig:corr_CMFT}(b), we plot two-site correlation functions for single, pair and three particles 
of the form $C_{1p}=\langle a_i^\dagger a_j\rangle$(red circles),  
$C_{2p}=\langle (a_i^\dagger)^2 (a_j)^2 \rangle$(green diamond) and $C_{3p}=\langle (a_i^\dagger)^3 (a_j)^3\rangle$(blue triangles). 
It can be seen that $C_{2p}$ is significantly larger than $C_{1p}$ and $C_{3p}$ which confirms 
the existence of the PSF phase. We identify the order of the phase transitions across the MI-ASF boundaries 
from the $\rho$ vs. $\mu$ behavior and from the hysteresis curves and mark them as dashed(dot-dashed) lines for first order 
(continuous) phase transitions in Fig.~\ref{fig:PD_CMFT}. 
The solid green circles correspond to the tricritical points. We also analyse the behaviour of the 
ground state energy with respect to the 
superfluid order parameter to verify the order of phase transitions(see supplementary material for details). 
Within our CMFT approach considered for model(~\ref{eq:zero}), 
we could not clearly determine the order of MI(1)-PSF-MI(3) and the PSF-ASF 
transitions. However, a rigorous analysis of the order parameters, the ground state energy or a second order 
hopping term in the Hamiltonian is needed to understand these phase 
transitions which will be studied else where. 
We would like to mention that 
the CMFT results for the ASF-MI phase boundaries match fairly well with the QMC results in Ref.~\cite{Sansone} 
in the strong interaction limit. 
We verify this by performing finite size extrapolation with the 
data for $2$, $4$ and $6$ sites clusters (see supplementary material details). 

A simple strong coupling analysis shows that for $J\ll U,W$, pair particles or paired holes move with an effective hopping 
 amplitude $\frac{6 J^2}{U+W}$. Hence, contrary to the mean field analysis, for any $0<J\ll U,W$ between the MI(1) and MI(3) 
lobes, an intermediate region of superfluid pairs may be found. For small $0<J\ll U,W$ the distance between MI(1) and MI(3) boundaries 
increases with $\sim J^2/(U-W)$ as is clearly observed in Figs.\ref{fig:PD_DMRG} and \ref{fig:PD_CMFT}. 
The single site mean-field theory approach~\cite{Sheshadri,Stoof} typically does not capture the PSF phase in such a model.
For a single decoupled site, the density of 
the system becomes $\rho=3$ after a pair of particles is added to the MI(1) phase due to the strong three-body attraction, avoiding the 
double occupancy. 
This seems like a first order MI(1)-MI(3) transition. As the CMFT method treats a multi-site system exactly, it successfully captures 
the PSF phase in between the Mott lobes.

In the following we discuss about the possible experimental observation of the PSF phase in the presence of an external confining 
potential. Before going to the inhomogeneous case, we would like to mention that the results discussed above for the model given in 
Eq.\ref{eq:zero} can be achieved by starting from the conventional form of the three-body interaction 
$\frac{W}{6}\sum_i n_i(n_i-1)(n_i-2)$ and including a four-body interaction $\frac{Q}{24}\sum_i n_i(n_i-1)(n_i-2)(n_i-3)$ which has been 
observed in the experiment~\cite{SWill}. It can be easily shown that in the limit $Q=-4W$ and at $\rho=3$, the system is equivalent 
to the one considered in Eq.~\ref{eq:zero}. Hence, by suitably engineering the two-, three- and four-body interaction, i.e. by making 
$W=-1.5U$ and $Q=-4W$ one can achieve the PSF phase which is discussed above. In a recent proposal it was shown that it is possible 
to independently control these three interactions in an optical lattice~\cite{Petrov1}. The choice of multi-body interactions i.e. 
$W=-1.5U$ and $Q=-4W$ which translates the system into the one considered in Eq.~\ref{eq:zero} is one of many possible combinations 
shown in Ref.~\cite{Petrov2}. Therefore, it can be made possible to achieve the PSF phase in the optical lattice experiment with the 
tunable multi-body interactions.

\begin{figure}[!t]
   \centering
\vspace{-0.85cm}
\includegraphics[trim={0.0cm 0.0cm 0cm 2.0cm},clip,width=0.35\textwidth,angle=-90]{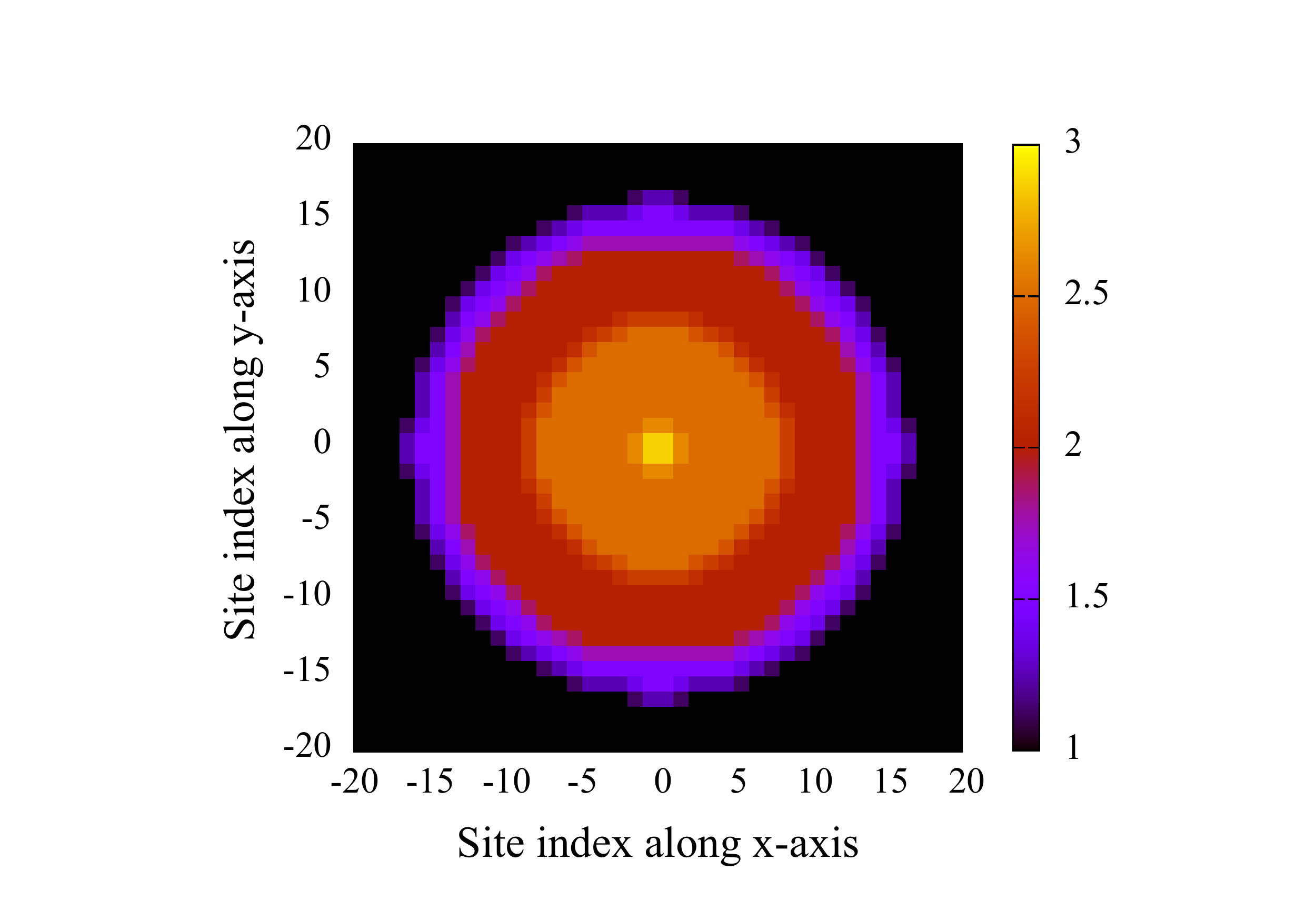}
\vspace{-0.4cm}
\caption{(Color online) MI(1), PSF and MI(3) phases in the presence of external confining potential $V_{trap}=0.002$ and 
at $zt/U=0.05$, $\mu/U=0.753125$.}
\label{fig:trap_fig}
\vspace{-0.3cm}
\end{figure}

Now we move on to the inhomogeneous case in the presence of external confining potential. As the PSF phase is very narrow compared to 
the MI lobes, one can consider a suitable trapping potential to see the PSF region. In our calculation we consider $V_{trap}=0.002$ and 
take a cut through the phase diagram corresponding to $zt/U=0.05$. Keeping the trap center just inside the MI(3) phase by fixing 
$\mu/U=0.753125$ 
we obtain the density profile as shown in Fig.\ref{fig:trap_fig}. A clear wedding cake structure depicting the MI plateaus and the 
intermediate PSF phase is obtained which we map and show here in two dimension for clarity. The black region is the MI(1) phase and the 
central bright square is the MI(3) phase. The region in between these two MI phases is the PSF phase. 
This PSF phase can possibly be detected by using the parity 
measurement discussed above using the single site imaging spectroscopy~\cite{Bakr,Endres}.


In summary, we re-investigate a Bose-Hubbard model in the presence of repulsive two-body ($U$) and attractive three-body ($W$) 
interaction where the three-body interaction is finite when the on-site particle number is three. By analyzing the one- and 
two-dimensional systems using the DMRG and CMFT approach respectively, we obtain the complete phase diagram and show that there exists 
a PSF phase in between the MI(1) and MI(3) lobes for $W=-1.5U$. This result is in contrast to the previous study where a direct first order 
phase transition from MI(1) to MI(3) has been reported in a two dimensional system using the single site mean-field 
theory and the QMC methods~\cite{Sansone}. By analyzing the finite size scaling of important physical quantities,   
we show that the PSF-ASF transition in the $1d$ case is 
of Ising universality class, whereas the MI(1)-PSF and PSF-MI(3) transitions are continuous. 
Within the CMFT approach we could not convincingly predict the ASF-PSF phase transitions. Although there exists a tricritical point along 
the MI(1)-ASF and ASF-MI(3) boundaries in the $2d$ phase diagram , there are no such points in the $1d$ case. 
We also discuss a scenario to achieve the model considered here by starting from the conventional forms of the multi-body 
interactions which is more suitable for ultracold atoms in optical lattice set up. In the end we discuss about the experimental 
realization of the PSF phase. 

\textit{Acknowledgment:-}
We would like to acknowledge Luis Santos for stimulating discussions. The computational simulations were carried out using 
the Param-Ishan HPC facility at Indian Institute of Technology - Guwahati, India and the Leibniz University of Hannover, Germany, computing 
facility. M.S. would like to acknowledge DST-SERB, India for the financial support through Project No.PDF/2016/000569. 
S.G. acknowledges support from the German Research Foundation DFG (project no. SA 1031/10-1).
T.M. acknowledges all the support from the start-up grant received from Indian Institute of Technology - Guwahati, India.

\end{document}


\title{Supplementary material for \textquotedblleft Anomalous pairing of bosons: Effect of multibody interactions in optical lattice\textquotedblright}
\author{Manpreet Singh$^{1}$, Sebastian Greschner$^{2,3}$ and Tapan Mishra$^{1}$}
\affiliation{$^{1}$Department of Physics, Indian Institute of Technology, Guwahati, Assam - 781039, India}
\affiliation{$^{2}$Institut f\"ur Theoretische Physik, Leibniz Universit\"at Hannover, 30167~Hannover, Germany}
\affiliation{$^{3}$Department of Quantum Matter Physics, University of Geneva, 1211 Geneva, Switzerland}
\date{\today}

\begin{abstract}
In this Supplementary Material we discuss additional details about the fidelity susceptibility calculations in $1d$, 
scaling of the CMFT results and comparison with QMC data and the nature of phase transitions in our model.
\end{abstract}

\maketitle

\section{Fidelity susceptibility}
\label{sec:fid}

To further complement the nature of ASF-PSF phase transition in $1d$, which is found to be of Ising universality class, 
we compute another quantity of interest called the fidelity susceptibility~\cite{Gu} which is 
given by 
\begin{align}
\chi_{FS}(t) = \lim_{\delta t\to 0} \frac{-2 \ln |\langle \Psi_0(t) | \Psi_0(t + \delta t) \rangle| }{(\delta t)^2} \,,
\end{align}
with $|\Psi_0 (t)\rangle$ being the ground-state wavefunction for a tunneling $t$ and $\delta t$ is a small parameter.
\begin{figure}[!h]
\centering
\includegraphics[width=0.9\linewidth]{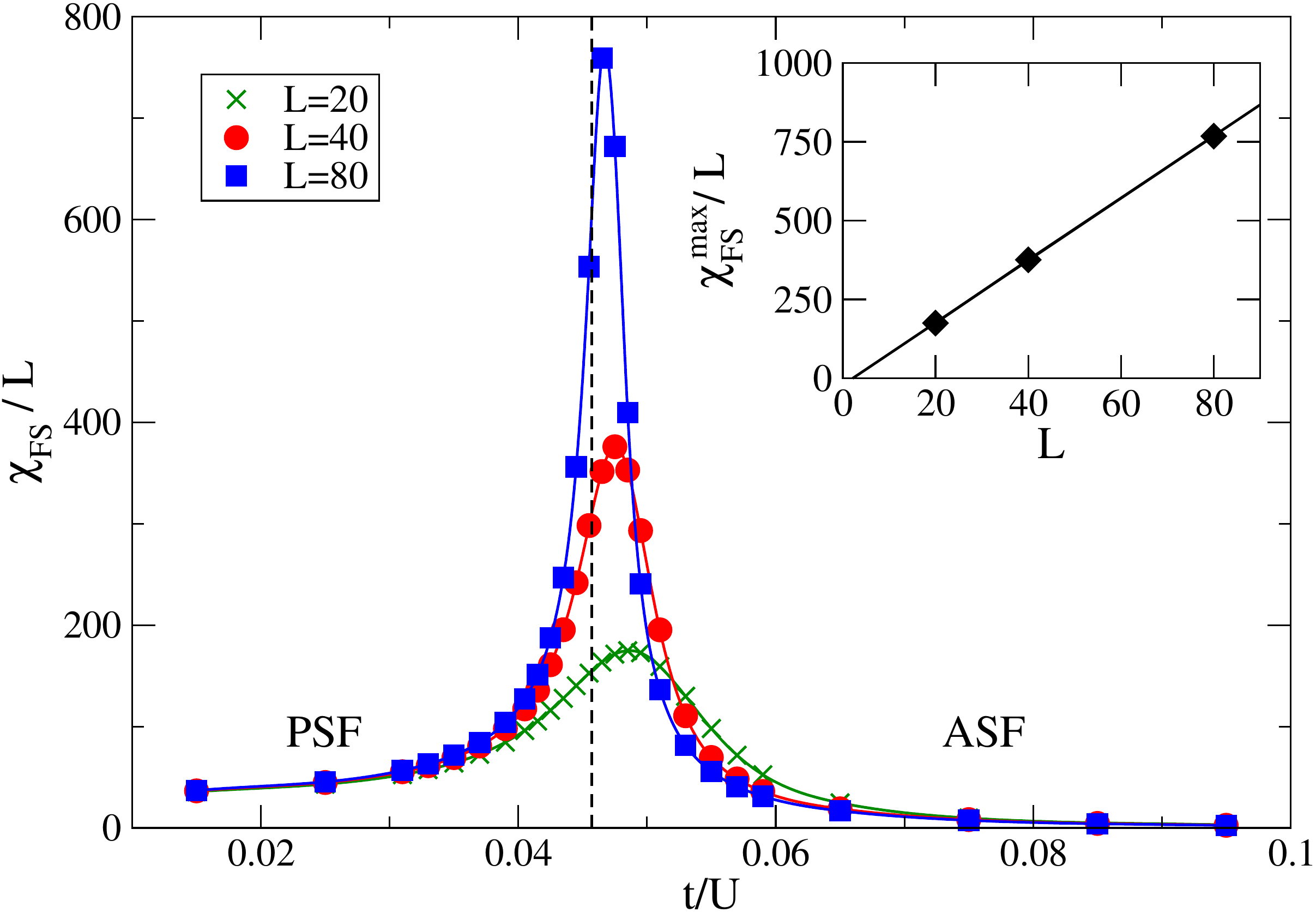}
\caption{Scaling of the fidelity susceptibility $\chi_{FS} /L $ as function 
of $t/U$. The inset shows the linear divergence of the 
peak height of the $\chi_{FS} /L $ curve with $L$.}
\label{fig:pofs}
\end{figure}
In Fig.~\ref{fig:pofs} we plot $\chi_{FS}(t)/L$ for different values of $t/U$ across the PSF-ASF phase transition at $\rho=2$ for 
$L=20, 40, 80$ sites. As the system size increases, the $\chi_{F}/L$ develops a distinct divergent single peak at 
$t/U \approx 0.045$ which is the critical point of the PSF-ASF transition at $\rho=2$. 
The linear scaling of the peak-height ($\chi^{max}_F/L$) with respect to its wings in the inset of Fig.~\ref{fig:pofs} confirms 
the Ising-character of this phase transition~\cite{Greschner2013b,Azimi2014}. 

\section{Comparison with the QMC results}
\label{sec:cmft_QMC}
\begin{figure}[!t]
   \centering
\includegraphics[width=0.46\textwidth]{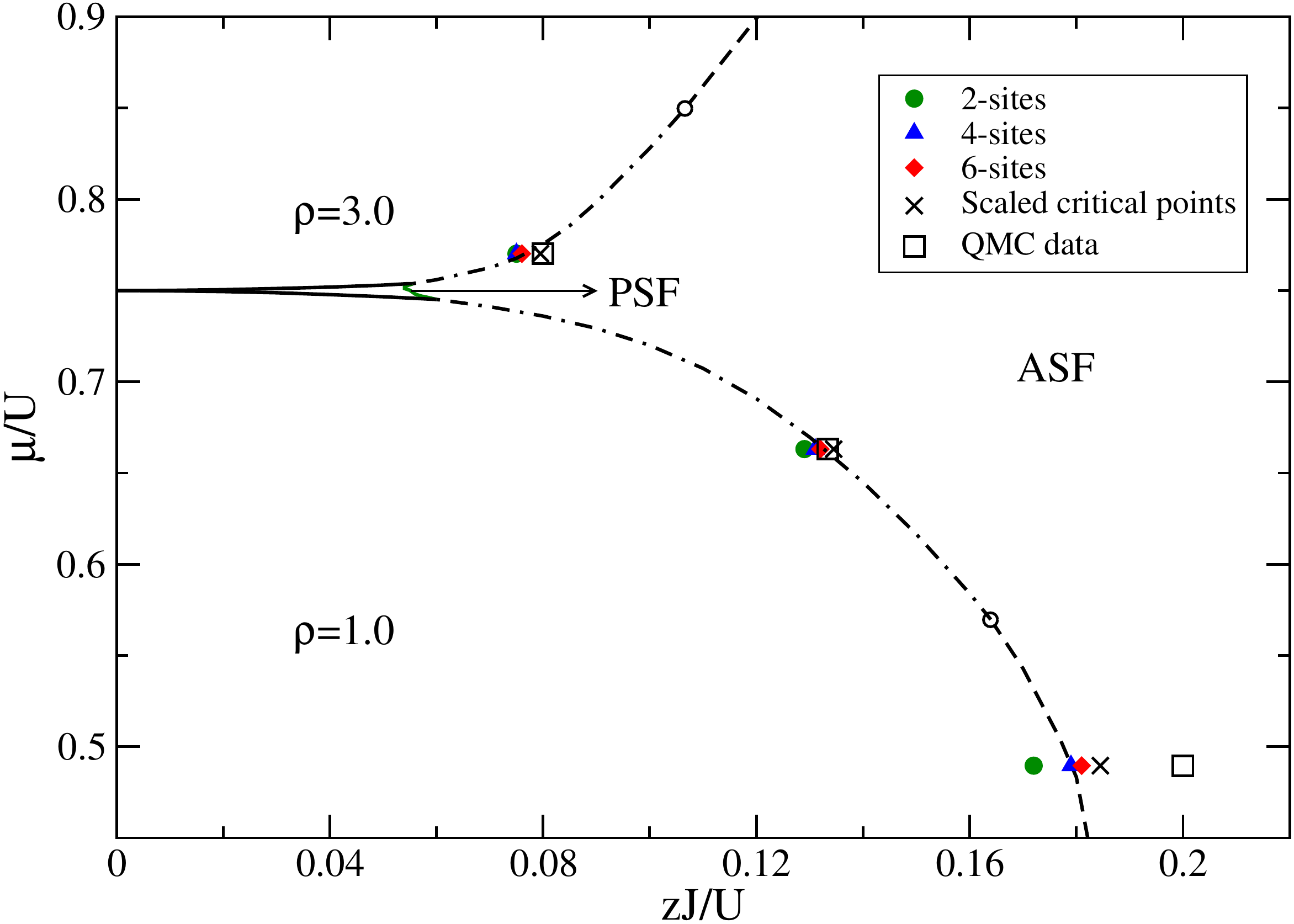}
\caption{(Color online) Comparison of 2-, 4- and 6-site CMFT results with QMC. The green circle (2-sites), blue triangle (4-sites)
and red diamond (6-sites) mark the $(zt/U)_{critical}$ for MI-ASF transition. The critical points obtained after
the scaling are marked with a $\times$ symbol. The squares represent the extracted QMC data from Fig.~3 of Ref.\cite{Sansone}}
\label{fig:cmft_QMC}
\end{figure}

In this section we show the comparison between the results obtained by our CMFT approach and the QMC results reported in Ref~\cite{Sansone} 
for a $2d$ system.
Considering cluster sizes of 2-, 4-, and 6-site we compute critical points across the MI-ASF boundaries by moving along the $zt/U$ axis for
fixed $\mu/U$ of the phase diagram shown in Fig. 4 of the main text. 
The values of $\mu/U$ selected are such that they match closely with the QMC data points given in Fig.3 of Ref.\cite{Sansone}. 
The finite size extrapolated points (denoted as symbol $\times$) obtained from the CMFT calculation match 
fairly well with the QMC results as shown in Fig.\ref{fig:cmft_QMC}.

\section{MI-ASF transitions}

\begin{figure}[!t]
   \centering
\includegraphics[width=0.46\textwidth]{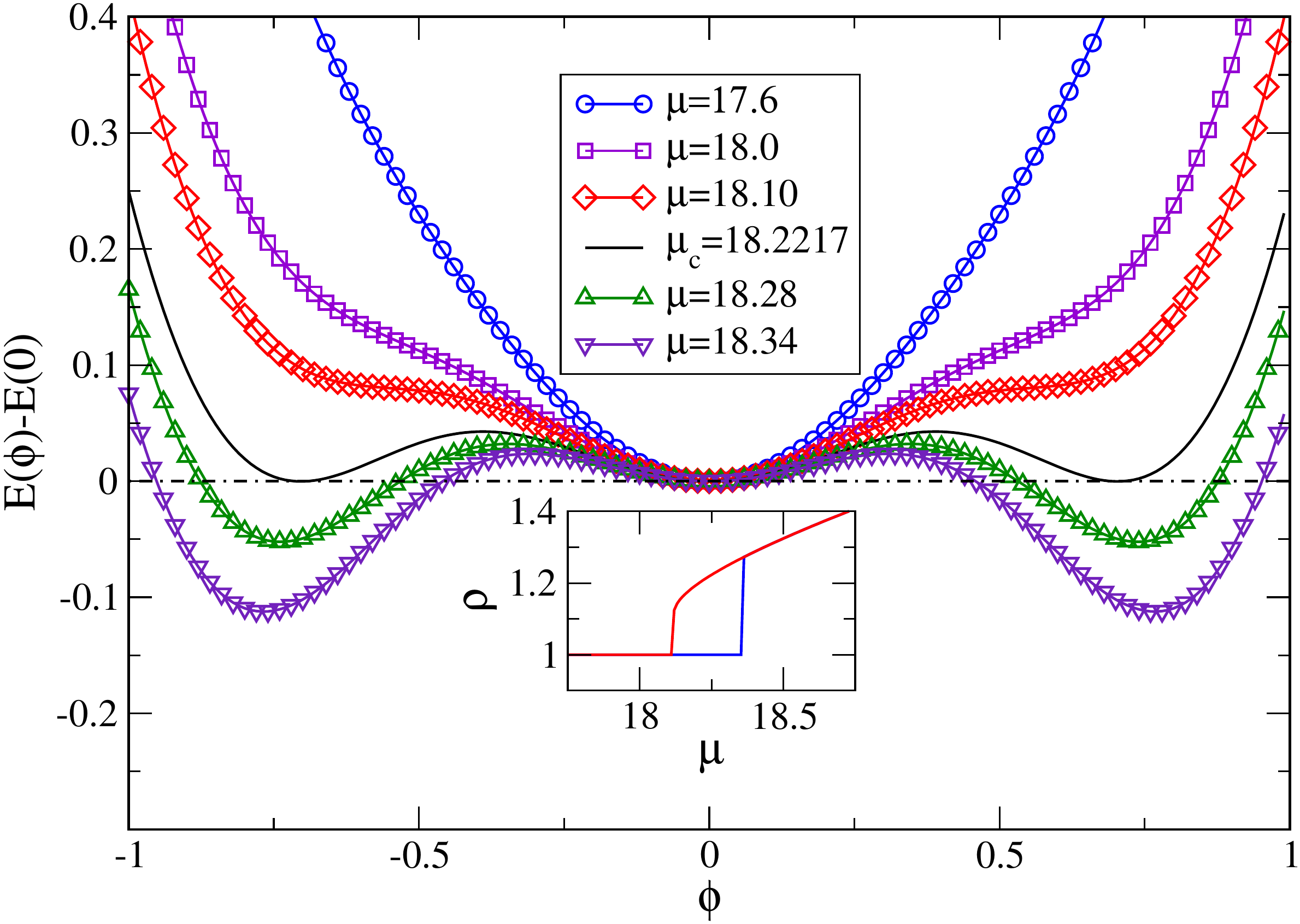}
\caption{(Color online) $E(\phi)-E(\phi=0)$ vs $\phi$ around the critical chemical potential across the MI-ASF phase boundary 
for $zt/U=0.14$.}
\label{fig:E-phi1}
\end{figure}

\begin{figure}[!t]
   \centering
\includegraphics[width=0.46\textwidth]{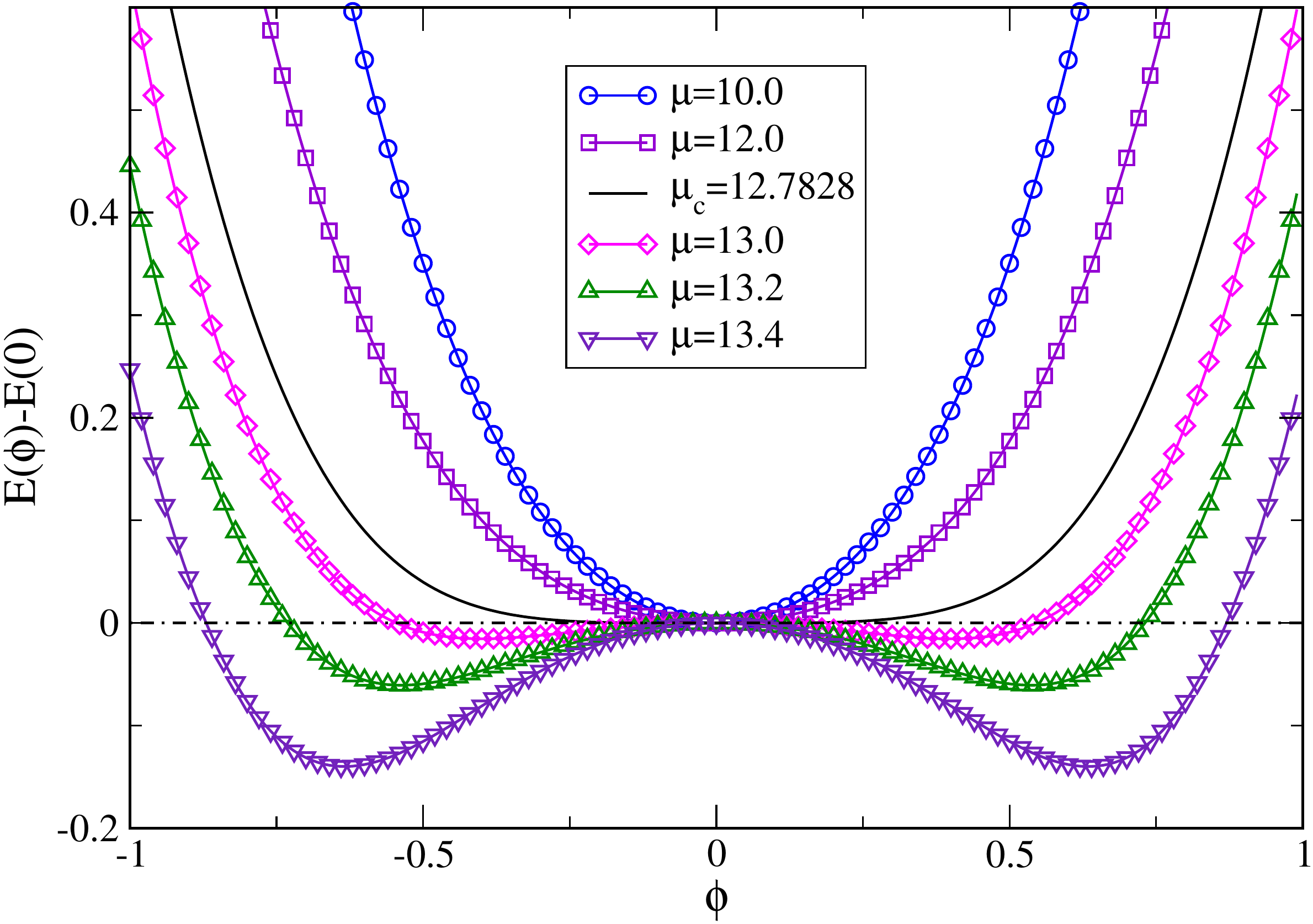}
\caption{(Color online) $E(\phi)-E(\phi=0)$ vs $\phi$ around the critical chemical potential across the MI-ASF phase 
boundary for $zt/U=0.17$}
\label{fig:E-phi2}
\end{figure}

In the following we show from our CMFT calculations that the MI-ASF transition boundaries constitute both first order as well as continuous 
regions. This can be understood by observing the ground state energy around the critical point~\cite{MSingh4}. 
We plot $E(\phi)-E(\phi=0)$ as a function of the superfluid order parameter $\phi$ close 
to the critical value of $\mu$ for the MI-ASF phase transition in Fig.~\ref{fig:E-phi1}.
We consider the value of $\mu=18.2217$ which corresponds to the MI(1)-ASF transition for $zt/U=0.14$. The appearance of three
degenerate minima at the critical $\mu$ indicates the first order transition. On the other hand the appearance of two 
degenerate minima from a single minimum in energy corresponds to a continuous MI(1)-ASF phase transition as shown in Fig.~\ref{fig:E-phi2},
where we consider the critical $\mu=12.7828$ at $zt/U=0.17$. The first order transition is further verified by the 
typical hysteresis behaviour of the $\rho$ vs $\mu$ curve~\cite{Sansone}as shown in the inset of Fig.~\ref{fig:E-phi1}. Similar 
feature is also seen in the ASF-MI(3) boundary. 

